\documentstyle[preprint,tighten,eqsecnum,aps,floats,psfig,epsfig]{revtex}

\def\ub{{\overline{u}}}
\def\vb{{\overline{v}}}

\def\e{\epsilon}

\begin{document}
\draft

\title{
Critical behavior of two-dimensional cubic and $MN$ models 
in the five-loop renormalization-group approximation} 

\author{P. Calabrese${}^{1}$, %\cite{PC-email}, 
E. V. Orlov${}^2$, D. V. Pakhnin${}^2$, and 
A. I. Sokolov$^{2}$,%\cite{AS-email}
}

\address{$^1$Rudolf Peierls Centre for Theoretical Physics, 1 Keble Road,
Oxford OX1 3NP, United Kingdom.}

\address{$^2$Department of Physical Electronics, Saint 
Petersburg Electrotechnical University, Professor Popov 
Street 5, St. Petersburg 197376, Russia.\\
{\bf e-mail: \rm 
{\tt calabres@thphys.ox.ac.uk},
{\tt ais2002@mail.ru},
}}
\date{\today}

\maketitle

\begin{abstract}
The critical thermodynamics of the two-dimensional $N$-vector cubic 
and $MN$ models is studied within the field-theoretical renormalization-group
(RG) approach. 
The $\beta$ functions and critical exponents are calculated in the five-loop 
approximation and the RG series obtained are resummed using the Borel-Leroy 
transformation combined with the generalized Pad\'e approximant and conformal 
mapping techniques. For the cubic model, the RG flows for various $N$ are 
investigated. For $N = 2$ it is found that the continuous line of fixed 
points running from the $XY$ fixed point to the Ising one is well reproduced 
by the resummed RG series and an account for the five-loop terms makes 
the lines of zeros of both $\beta$ functions closer to each another. 
For the cubic model with $N\geq 3$, the five-loop contributions 
are shown to shift the cubic fixed point, given by the four-loop 
approximation, towards the Ising fixed point. 
This confirms the idea that the existence of the cubic fixed point in two 
dimensions under $N > 2$ is an artifact of the perturbative analysis. 
For the quenched dilute $O(M)$ models ($MN$ models with $N=0$) the results 
are compatible with a stable pure fixed point for $M\geq1$.
For the $MN$ model with $M,N\geq2$ all the non-perturbative results are 
reproduced. In addition a new stable fixed point is found for moderate
values of $M$ and $N$.

\end{abstract}

\pacs{PACS Numbers: 75.10.Hk, 05.70.Jk, 64.60.Fr, 11.10.Kk}

\newpage

\section{Introduction}

The two-dimensional model with $N$-vector order parameter 
and cubic anisotropy is known to have a rich phase diagram; 
it contains, under different values of $N$ and of the anisotropy 
parameter, the Ising-like and Kosterlitz-Thouless critical 
points, lines of the first-order phase transitions, and the 
line of the second-order transitions with continuously varying 
critical exponents (see, e. g. \cite{JKKN,Sh94,CC01} for review). 
This model is related to many other familiar models in various 
particular cases, while for $N \to 0$ it describes the critical 
behavior of two-dimensional weakly disordered Ising systems. Moreover, 
exact solutions are known for the two-dimensional cubic model in the 
several limits such as an Ising decoupled limit, the limit of 
extremely strong anisotropy for $N > 2$ \cite{Sh94,Sh89} and 
the replica limit $N \to 0$ \cite{Sh94,DD83,Sh84}. 
The correspondence, in particular regions of the phase diagram, with the 
$N$-color Ashkin-Teller models, discrete cubic models, and planar model with
fourth order anisotropy give further informations about the critical 
behavior. 
All these issues are reviewed in Ref. \cite{CC01} and we will not repeat 
them here.
These features make the two-dimensional $N$-vector cubic model a convenient 
and, perhaps, unique testbed for evaluation of the analytical and numerical 
power of perturbative methods widely used nowadays in the 
theory of critical phenomena. The field-theoretical 
renormalization-group (RG) approach in physical dimensions 
is among of them. 

Recently, the critical behavior of the two-dimensional $N$-vector cubic 
model was explored using the renormalization-group technique 
in the space of fixed dimensionality \cite{CC01}. The four-loop 
expansions for the $\beta$-functions and critical exponents 
were calculated and analyzed using the Borel transformation 
combined with the conformal mapping and Pad\'e-approximant 
techniques as a tool for resummation of the divergent RG 
series. The most part of predictions obtained within the 
renormalization group approach turned out to be in accord 
with known exact results. At the same time, some findings 
were quite new. In particular, for $N > 2$ the resummed 
four-loop RG expansions for $\beta$-functions were found 
to yield a cubic fixed point with (almost) marginal 
stability; this point does not correspond to any of the 
critical asymptotics revealed by exact methods ever 
applied. Although the stability properties of the cubic 
fixed point look very similar to those of its Ising 
counterpart, these points were found to lie too far from 
each other (for moderate $N$) to consider the distance 
between them as a splitting caused by the limited accuracy 
of the RG approximation employed. 

It is worthy to note that this situation is quite different 
from what we have in three dimensions. Indeed, for the 
three-dimensional cubic model the structure of the RG flow diagram is 
known today with a rather high accuracy. Recent five-loop 
\cite{Klein,SAS,PS} and six-loop \cite{CPV,FHY} RG calculations 
certainly confirmed that for $N > 2$ the cubic fixed point 
does not merge with any other fixed point and governs the 
specific anisotropic mode of critical behavior, 
distinguishable from the Ising and Heisenberg modes 
(see, e. g. \cite{PS01}). 

It is very desirable, therefore, to clear up to what 
extent the location of the cubic fixed point in two dimensions 
is sensitive to the order of the RG calculations and, more 
generally, whether this point really exists at the flow diagram 
or its appearance is the approximation artifact caused by the 
finiteness of the perturbative series and by an ignorance of 
the confluent singularities significant in two dimensions 
\cite{Nick91,PV98,OS,CCCPV}.  

Of prime interest is also the situation with the line of 
fixed points that should run, under $N = 2$, from the Ising 
fixed point to the $XY$ one. Within the four-loop 
approximation, the zeros of $\beta$-functions for the 
$O(N)$-symmetric and anisotropic coupling constants form two 
lines that for $N = 2$ are practically parallel to each other 
and separated by the distance that is smaller than the error 
bar appropriate to the working approximation \cite{CC01}. 
Will the higher-order contributions keep these two lines parallel? 
Will an account for the higher-order terms further diminish the 
distance between these lines or their splitting should be 
attributed, at least partially, to the influence of the 
singular terms just mentioned?  
 
To answer the above questions, it is necessary to analyze 
the critical behavior of the two-dimensional cubic model in the higher 
perturbative orders. Recently, the renormalization-group 
expansions for the two-dimensional $O(N)$-symmetric model were obtained 
within the five-loop approximation \cite{OS}. In the course of 
this study, all the integrals corresponding to the five-loop 
four-leg and two-leg Feynman graphs have been evaluated. 
This makes it possible to investigate the critical 
thermodynamics of anisotropic two-dimensional models with several couplings 
in the five-loop approximation. In this paper, such an 
investigation will be carried out for the two-dimensional $N$-vector 
model with cubic anisotropy.

For studying the effect of cubic anisotropies one usually considers 
the $\phi^4$ theory \cite{Aharony-76,rev-01}:
\begin{eqnarray}
{\cal H} = \int d^d x &&
\left\{ {1\over 2} \sum_{i=1}^{N}
      \left[ (\partial_\mu \phi_i)^2 +  r \phi_i^2 \right]  +
 {1\over 4!} \sum_{i,j=1}^N \left( u_0 + v_0 \delta_{ij} \right)
\phi^2_i \phi^2_j \right\}, 
\label{Hphi4}
\end{eqnarray}
in which the added cubic term breaks explicitly the $O(N$) invariance 
leaving a residual discrete cubic symmetry given by the reflections
and permutations of the field components. 
This term favors the spin orientations towards the faces or the corners of
an $N$-dimensional hypercube for $v_0<0$ or $v_0>0$ respectively.
In two dimensions the effect of anisotropy is particularly important:
systems possessing continuous symmetry do not exhibit conventional long-range
order at finite temperature, while models with discrete symmetry do undergo
phase transitions into conventionally ordered phase.

In general, the model (\ref{Hphi4}) has four fixed points:
the trivial Gaussian one, the Ising one in which the $N$ components 
of the field decouple,
the O($N$)-symmetric and the  cubic fixed points.
The Gaussian fixed point is always unstable, and so is 
the Ising fixed point for $d>2$~\cite{Aharony-76}.
Indeed, in the latter case, it is natural to interpret Eq. (\ref{Hphi4}) 
as the Hamiltonian of $N$ Ising-like systems coupled by the 
$O(N)$-symmetric term. But this interaction is the sum of the products 
of the energy operators of the different Ising systems.
Therefore, at the Ising fixed point, the crossover exponent 
associated with the O($N$)-symmetric quartic term is given by
the specific-heat critical exponent $\alpha_I$ of the Ising model,
independently of $N$ (for $N=0$ this argument is equivalent to the 
Harris criterion \cite{Harris}). 
Since $\alpha_I$ is positive for all $d>2$ 
the Ising fixed point is unstable. Obviously, in two dimensions this argument
only tells us that the crossover exponent at this fixed point vanishes.
Higher order corrections to RG equations may lead either to a marginally 
stable fixed point \cite{gw-81} or to a line of fixed points.  
It was argued that for $N\geq 3$ the former possibility is realized, while 
for $N=2$ the latter one holds (see Ref. \cite{JKKN,Sh94,CC01} and 
references therein).

The stability properties of the fixed points depend on $N$.
For sufficiently small values of $N$, $N<N_c$, the 
O($N$)-symmetric fixed point is stable and the cubic one  is unstable.
For $N>N_c$ and $d>2$, the opposite is true:
the renormalization-group flow is driven towards the cubic fixed point, 
which now describes the generic critical behavior of the system.
At $N=N_c$, the two fixed points should coincide for $d>2$.
At $d=2$, it is expected that $N_c=2$ \cite{pn-76} and a line of fixed points
connecting the Ising and the $O(2)$-symmetric fixed points 
exists \cite{JKKN,CC01}.

A generalization of the Hamiltonian Eq. (\ref{Hphi4}) is obtained by
considering $N$ coupled $O(M)$ vector models, instead of $N$ Ising models.
The resulting Hamiltonian (defining the so called $MN$ model) 
is \cite{Aharony-76,rev-01}
\begin{eqnarray}
{\cal H}_{\rm MN} = \int d^d x 
\Bigl\{ \sum_{i,a}{1\over 2} 
\left[ (\partial_\mu \phi_{a,i})^2 + r \phi_{a,i}^2 \right]  
+
 \sum_{ij,ab} {1\over 4!}\left( u_0 + v_0 \delta_{ij} \right)
\phi^2_{a,i} \phi^2_{b,j} \Bigr\},
\label{Hphi4MN}
\end{eqnarray}
where $a,b=1,\dots M$ and $i,j=1,\dots N$.
The continuous $O(MN)$ symmetry is explicitly broken by the $v_0$ term to
$C_N\times O(M)$ where $C_N$ is the discrete group of permutations of 
$N$ elements. The presence of such discrete symmetry allows for a 
finite temperature phase transition under general values of $N$ 
and $M$ (differently from models with $v_0=0$).
For $M=1$ it reduces to the Hamiltonian of the cubic model,
but it has physical applications for $M\neq1$ too.
For $N\rightarrow0$ and generic $M$, under the condition $u_0<0$ and $v_0>0$,
it describes the critical behavior of quenched dilute 
$O(M)$ models \cite{rev-01,revran}.
For $M=2$ and $N=2$ ($N=3$) and $v_0>0$, it is relevant for the second-order 
phase transition in planar (isotropic) antiferromagnets with complicated
ordering as the three dimensional sinusoidal magnets 
TbAu$_2$, DyC$_2$, type-II antiferromagnets TbAs, TbP (type-III 
antiferromagnets K$_2$IrCl$_6$, sinusoidal TbD$_2$) and many 
others \cite{mk-76}.
For $M=N=2$ and $v_0<0$ an exact mapping \cite{AS-94} relates the Hamiltonian
(\ref{Hphi4MN}) with the $O(2)\times O(2)$ symmetric one, describing the 
critical behavior of frustrated antiferromagnets with non-collinear 
order \cite{revchirali,rev-01}.
For $N\rightarrow\infty$ it describes $O(M)$ models with 
constrains \cite{emery-75} (as $O(M)$ models with annealed disorder).

The Hamiltonian (\ref{Hphi4MN}) has been largely studied in the framework 
of $\e$ expansion ($\e=4-d$) \cite{epsMN,cp-prep} and directly in three 
dimensions \cite{cp-prep,3dMN,pv-00,dhy-04,HDY-01},
mainly to understand the critical behavior of the 
antiferromagnets quoted above. 
Much less studies were devoted to the two dimensional case for $M\neq1$,
since its main features can be understood by non-perturbative arguments. 
In fact, three fixed points always exist in the RG flow for any $N$ and $M$: 
the always unstable Gaussian ($u_0=v_0=0$), the $O(M)$ (with $u_0=0$) and 
the $O(NM)$ (with $v_0=0$).
Their stability properties are known from exact arguments.
As for the Ising fixed point in the cubic model, 
the stability of the $O(M)$ fixed point is governed by the specific-heat 
critical exponent $\alpha_M$ of the $O(M)$ model. 
Being (in two dimensions) $\alpha_M$ negative for all $M>1$, 
the $O(M)$ fixed point is stable for physical values of $M$.
An exact argument ensures that the $O(NM)$ fixed point is unstable 
when $NM>N_c=2$ \cite{pn-76,cpvmult}, thus it is always unstable for physical 
values (apart from the replica limit $N\rightarrow 0$, but in this case it is
not reachable from the physical initial condition with $u_0>0$).
In the framework of $\e$ expansion another fixed point (called mixed) 
exists \cite{Aharony-76}. 
It is the generalization to $M\neq1$ of the cubic one.
Its analytic continuation to $d=2$ is expected to be 
in the region $u_0<0$ and $v_0>0$ for $N,M\geq2$ and
in the $u_0,v_0>0$ region (unphysical) in the replica limit. 
It is expected to be unstable in both the cases.

Anyway, these arguments still do not completely characterize the RG flow of
the $MN$-model. 
Indeed for $N,M\geq2$ and $v_0<0$ the situation is more controversial. 
The $\e$ expansion indicates that no fixed point exists and
consequently it was longly believed that no continuous transition can 
take place in these systems.
Nevertheless, for $N=M=2$ the quoted mapping between the $MN$ and the 
$O(2)\times O(2)$ models shows the presence of a fixed point (found in the 
$O(2)\times O(2)$ model at four \cite{cp-01} and 
five loops \cite{cps-02,cops-02}) in the region $v_0<0$. 
Consequently the transition should be second order in the chiral
universality class.
The natural question arising is whether the fixed point found at $N=M=2$ is a 
peculiar feature or it persists to some larger values of $N$ and $M$~(not
too large, since for $N\gg 1$, at fixed $M$, we expect that the $\e$ 
expansion can be smoothly continued up to $d=2$).
An answer to this question can be found within higher order perturbative 
expansion.

The paper is organized as follows. 
In Section \ref{sec2} the five-loop 
contributions to the renormalization-group functions are  
calculated and the singularities of Borel transforms of the 
renormalization-group series are discussed for the general $MN$ model.
Section \ref{sec3} is devoted to the analysis of the critical behavior of the 
model in the five-loop approximation. 
The existence of a cubic fixed point for $N\geq3$ is investigated.
The critical behavior of quenched random $O(M)$ models is considered.
Attention is paid to the reproducibility of a continuous line of fixed 
points, predicted earlier for the planar ($N=2,\,M=1$) model, in the 
framework of the field-theoretical RG approach. 
The critical exponents along this line are evaluated in this section as well. 
Finally, the critical behavior for $N,M\geq 2$ is addressed.
In Section \ref{sec4} we summarize the main results obtained and make 
concluding remarks.

\section{Fixed dimension perturbative expansions}
\label{sec2}

\subsection{Renormalization of the theory}
\label{sec2a}

The fixed-dimension field-theoretical approach represents a 
powerful procedure in the study of the critical properties 
of three-dimensional systems belonging to the $O(N)$ and more complicated 
universality classes (see, e.g., Ref.~\cite{rev-01,ZJ-book,cprv-02}).  
In this approach one performs an expansion in powers of appropriately defined 
zero-momentum quartic
couplings and renormalizes the theory by a set of zero-momentum conditions
for the (one-particle irreducible) two-point and four-point correlation 
functions. For the $MN$ model they read:
\begin{equation}
\Gamma^{(2)}_{ab,ij}(p) = \delta_{ai,bj} Z_\phi^{-1} \left[ m^2+p^2+O(p^4)\right],
\label{ren1}  
\end{equation}
where $\delta_{ai,bj}=\delta_{ab}\delta_{ij}$, and 
\begin{eqnarray}
\Gamma^{(4)}_{abcd,ijkl}(0) =& 
Z_\phi^{-2} m^2 &\left[  
\frac{u}{3}\left(\delta_{ai,bj}\delta_{ck,dl} +\delta_{ai,ck}\delta_{bj,dl} + 
                \delta_{ai,dl}\delta_{bj,ck} \right)+
\right. \nonumber\\&&\left.
\frac{v}{3} \,\delta_{ij}\delta_{ik}\delta_{il}
(\delta_{ab}\delta_{cd}+\delta_{ac}\delta_{bd}+\delta_{ad}\delta_{bc})\right].
\label{ren2}  
\end{eqnarray}
They relate the inverse correlation length (mass) $m$ and the zero-momentum
quartic couplings $u$ and $v$ to the corresponding Hamiltonian parameters
$r$, $u_0$, and $v_0$:
\begin{equation}
u_0 = m^2 u Z_u Z_\phi^{-2},\qquad\qquad
v_0 = m^2 v Z_v Z_\phi^{-2}.
\end{equation}
In addition, one introduces the function $Z_t$ defined by the relation
\begin{equation}
\Gamma^{(1,2)}_{ai,bj}(0) = \delta_{ai,bj} Z_t^{-1},
\label{ren3}
\end{equation}
where $\Gamma^{(1,2)}$ is the (one-particle irreducible)
two-point function with an insertion of $\case{1}{2}\phi^2$.

From the perturbative expansion of the correlation functions
$\Gamma^{(2)}$, $\Gamma^{(4)}$, and $\Gamma^{(1,2)}$ and 
the above relations, one derives the functions $Z_\phi(u,v)$, 
$Z_u(u,v)$, $Z_v(u,v)$, and $Z_t(u,v)$ as double expansions in $u$ and $v$.

The fixed points of the theory are given by 
the common zeros of the $\beta$-functions
\begin{equation}
\beta_u(u,v) = m \left. {\partial u\over \partial m}\right|_{u_0,v_0} ,
\qquad\qquad
\beta_v(u,v) = m \left. {\partial v\over \partial m}\right|_{u_0,v_0} .
\label{bet}
\end{equation}
The stability properties of the fixed points are controlled  by the 
eigenvalues $\omega_i$ of the matrix 
\begin{equation}
\Omega = \left(\matrix{\displaystyle \frac{\partial \beta_u(u,v)}{\partial u}
 &\displaystyle \frac{\partial \beta_u(u,v)}{\partial v}
\cr & 
 \cr \displaystyle \frac{\partial \beta_v(u,v)}{\partial u}
& \displaystyle  \frac{\partial \beta_v(u,v)}{\partial v}}\right)\; ,
\end{equation}
computed at the given fixed point:
a fixed point is stable if both eigenvalues are positive.
The eigenvalues $\omega_i$ are related to
the leading scaling corrections, which vanish as
$\xi^{-\omega_i}\sim |t|^{\Delta_i}$ where $\Delta_i=\nu\omega_i$.
If $\omega_i$ has a non-vanishing imaginary part, the approaching to the FP
is spiral-like and the FP is called focus \cite{cps-02}.

One also introduces the functions
\begin{eqnarray}
\eta_{\phi,t}(u,v) &=& \left. {\partial \ln Z_{\phi,t} \over \partial \ln m}
         \right|_{u_0,v_0}
= \beta_u {\partial \ln Z_{\phi,t} \over \partial u} +
\beta_v {\partial \ln Z_{\phi,t} \over \partial v} , \label{eta12}
%\\ \eta_t(u,v) &=& \left. {\partial \ln Z_t \over \partial \ln m}
%         \right|_{u_0,v_0}
%= \beta_u {\partial \ln Z_t \over \partial u} +
%\beta_v {\partial \ln Z_t \over \partial v}.\label{eta2}
\end{eqnarray}
in terms of that, the critical exponents are obtained as
\begin{eqnarray}
\eta &=& \eta_\phi(u^*,v^*),
\label{eta_fromtheseries} \\
\nu &=& \left[ 2 - \eta_\phi(u^*,v^*) + \eta_t(u^*,v^*)\right] ^{-1},
\label{nu_fromtheseries} \\
\gamma &=& \nu (2 - \eta),
\label{gamma_fromtheseries} 
\end{eqnarray}
where $(u^*,v^*)$ is the location of the stable fixed point.

\subsection{The five loop series}
\label{sec2b}

We calculate the two-dimensional perturbative RG functions of the $MN$ model 
in the zero-momentum massive renormalization approach introduced before up to 
five loops. Corresponding contributions to the functions Eqs. (\ref{ren1}), 
(\ref{ren2}), and (\ref{ren3}) are given by 162 diagrams for the four-point 
correlators and by 26 graphs for the two-point one~\cite{KS-01}.
We handle them with a symbolic manipulation program, which  
generates the diagrams and computes the symmetry and group factors of 
each of them.
The RG functions are written in terms of the rescaled couplings
\begin{equation}
u \equiv  {8 \pi\over 3} \; R_{NM} \; \ub,\qquad\qquad
v \equiv   {8 \pi\over 3} \; R_{M} \vb ,
\label{resc}
\end{equation}
where $R_K = 9/(8+K)$.

The obtained series are
\begin{equation}
\bar{\beta}_{\ub}  = -\ub + \ub^2 + {2(M+2)\over (M+8)} \ub \,\vb+ 
\ub \sum_{i+j\geq 2} b^{(u)}_{ij} \ub^i \vb^j, 
 \label{bu}
\end{equation}
\begin{equation}
\bar{\beta}_{\vb}  = -\vb + \vb^2 + {12\over 8+NM} \ub \,\vb+
\vb \sum_{i+j\geq 2} b^{(v)}_{ij} \ub^i \vb^j,  \label{bv} 
\end{equation}
\begin{equation}
\eta_\phi = \sum_{i+j\geq 2} e^{(\phi)}_{ij} \ub^i \vb^j,
\label{etaphi}
\end{equation}
\begin{equation}
\eta_t = -{2 (2+N M)\over (8+N M) } \ub -{2+M\over 8+M} \vb +
\sum_{i+j\geq 2} e^{(t)}_{ij} \ub^i \vb^j
\label{etat}
\end{equation}
where
\begin{equation}
\bar{\beta}_{\ub}= {3\over 16\pi} \, R_{N M}^{-1}\beta_{u},\qquad\qquad
 \bar{\beta}_{\vb}= {3\over 16\pi} \, R_{M}^{-1} \beta_{v}.
\label{resc2}
\end{equation}
The coefficients  $b^{(u)}_{ij}$,  $b^{(v)}_{ij}$, $e^{(\phi)}_{ij}$, and 
$e^{(t)}_{ij}$  are reported in the Tables \ref{t1}, \ref{t2}, \ref{t3}, and 
\ref{t4}.
Note that due to the  rescaling (\ref{resc2}) and (\ref{resc}), 
the matrix element of  $\Omega$ 
are the double of the derivative of $\bar{\beta}$ with respect to $\ub$ 
and $\vb$.

We have verified the exactness of our series by the following relations:

(i) $\bar{\beta}_{\ub}(\ub,0)$, $\eta_\phi(\ub,0)$ 
and $\eta_t(\ub,0)$ reproduce
the corresponding functions of the $O(NM)$-symmetric
model~\cite{OS,OS-corr}.

(ii) $\bar{\beta}_{\vb}(0,\vb)$, $\eta_\phi(0,\vb)$ and 
$\eta_t(0,\vb)$ reproduce
the corresponding functions of the $O(M)$-symmetric $\phi^4$ 
theory \cite{OS,OS-corr}.

(iii) The following relations hold close to Heisenberg 
$O(NM)$ \cite{cpv-03,cppv-03}
\begin{eqnarray}
\left. {\partial \eta_{\phi,t}\over \partial \ub} \right|_{(\ub,0)}&=&
\frac{M N+2}{M+2}\frac{M+8}{MN+8}
\left. {\partial \eta_{\phi,t}\over \partial \vb} \right|_{(\ub,0)},
\label{euevrel}\\
\left. {\partial \bar{\beta}_\ub\over\partial \ub}\right|_{(\ub,0)}&-&
\left. {\partial \bar{\beta}_\vb\over\partial \vb}\right|_{(\ub,0)}=
\frac{M N+2}{M+2}\frac{M+8}{MN+8} 
\left. {\partial \bar{\beta}_\ub\over\partial \vb}\right|_{(\ub,0)},
\end{eqnarray}
and to the $O(M)$  fixed points \cite{cppv-03}
\begin{eqnarray}
\left. {\partial \bar{\beta}_\vb \over\partial \vb}\right|_{(0,\vb)}-
\left. {\partial \bar{\beta}_\ub \over\partial \ub}\right|_{(0,\vb)}&=& 
\frac{MN+8}{M+8} 
\left. {\partial \bar{\beta}_\vb \over\partial \ub}\right|_{(0,\vb)},\\
\left. {\partial \eta_\phi \over\partial \vb}\right|_{(0,\vb)}&=&
\frac{MN+8}{M+8} 
\left. {\partial \eta_\phi \over\partial \ub}\right|_{(0,\vb)}.
\end{eqnarray}

(iv) The following relations hold for $N=1$ and generic $M$: 
\begin{eqnarray}
&& \bar{\beta}_{\ub}(\ub,x-\ub) + \bar{\beta}_{\vb}(\ub,x-\ub) =\bar{\beta}_{\vb}(0,x),\nonumber\\
&&\eta_\phi(\ub,x-\ub) = \eta_\phi(0,x),\nonumber\\
&&\eta_t(\ub,x-\ub) = \eta_t(0,x).
\end{eqnarray}

(v) For $N=2$ and $M=1$, one easily obtains the identities~\cite{Korz-76,CC01}
\begin{eqnarray}
&& \bar{\beta}_{\ub}(\ub+\case{5}{3}\vb,-\vb) + 
{5\over 3}\bar{\beta}_{\vb}(\ub+\case{5}{3}\vb,-\vb) = 
\bar{\beta}_{\ub}(\ub,\vb),\\
&& \bar{\beta}_{\vb}(\ub+\case{5}{3}\vb,-\vb) = -\bar{\beta}_{\vb}(\ub,\vb), \nonumber\\
&&\eta_\phi(\ub+\case{5}{3}\vb,-\vb) = \eta_\phi(\ub,\vb),\nonumber\\
&&\eta_t(\ub+\case{5}{3}\vb,-\vb) = \eta_t(\ub,\vb).\nonumber
\end{eqnarray}
These relations are also exactly satisfied by our five-loop series.
Note that, since the Ising fixed point is $(0,g^*_I)$, and $g^*_I$ is known 
with high precision \cite{chpv-00}
\begin{equation}
g^*_I=1.7543637(25), \label{gI*}
\end{equation}
the above symmetry gives us 
the location of the cubic fixed point: $(\case{5}{3} g^*_I,-g^*_I)$.

(vi) In the large-$N$ limit the critical exponents of the mixed fixed point 
are related to those of the $O(M)$ model:
$\eta=\eta_M$ and $\nu = \nu_M$ \cite{emery-75,Ah-73}~(note that in
general the latter equivalence holds only for $M\geq 1$ and $d=2$, in the 
general case Fisher renormalization of exponents \cite{fisher-68} has to be 
taken into account \cite{Ah-73}).
One can easily see that, for $N\to\infty$,
$\eta_\phi(u,v) = \eta_\phi(0,v)$, where $\eta_\phi(0,v)$ is the perturbative
series that determines the exponent $\eta_M$ of the $O(M)$ model.
Therefore, the first relation is trivially true. On the other hand,
the second relation $\nu = \nu_M$ is not identically
satisfied by the series, and is verified only at the critical 
point \cite{CPV}.

(vii) For $M=0$, the $N$-independent series satisfy \cite{K83}
\begin{eqnarray}
\bar{\beta}_\ub(\ub,x-\ub)&+&\bar{\beta}_\vb(\ub,x-\ub)=\bar{\beta}_\ub(x,0)\,,\nonumber\\
\eta_\phi(\ub,x-\ub)&=&\eta_\phi(x,0)\,,\nonumber\\
\eta_t(\ub,x-\ub)&=&\eta_t(x,0)\,.
\end{eqnarray}

(viii) For $M\rightarrow\infty$ the series, as expected, reduce to
\begin{eqnarray}
\bar{\beta}_\ub=-\ub+\ub^2+2\ub\,\vb,&\qquad&\bar{\beta}_\vb=-\vb+\vb^2,\nonumber\\
\eta_t=-2\ub-\vb,&\qquad&\eta_\phi=0,
\end{eqnarray}
with two Gaussian and two spherical (i.e. $O(\infty)$) fixed points (as the 
$O(N)\times O(M)$ model \cite{prv-01n}).

(ix) For $N=M=2$ they agree, according to the exact mapping \cite{AS-94}:
\begin{equation} 
\ub=\ub_{\rm ch}+\frac{\vb_{\rm ch}}{2}\,,\qquad
\vb=-\frac{5}{6}\vb_{\rm ch}\,,
\end{equation}
with the five-loop series of the $O(2)\times O(2)$ model \cite{cops-02},
written in terms of $\ub_{\rm ch}$ and $\vb_{\rm ch}$.

(x) The series reproduce the previous four-loop results 
for $M=1$ \cite{CC01}.

%betau 
\begin{table}[tbp]
\squeezetable
\caption{
The coefficients $b^{(u)}_{ij}$, cf. Eq. (\ref{bu}).
}
\label{t1}
\begin{tabular}{cl}
\multicolumn{1}{c}{$i,j$}&
\multicolumn{1}{c}{$R_{NM}^{-i} R_M^{-j} b^{(u)}_{ij}$}\\
\tableline \hline
2,0&$-0.588581 - 0.127593\,M\,N$\\
1,1&$-0.621503 - 0.310751\,M$\\
0,2&$-0.144054 - 0.0720268\,M$\\
3,0&$ 0.719309 + 0.204598\,M\,N + 0.00685909\,M^2\,N^2$\\
2,1&$ 1.19188 + 0.595938 M + 0.0598976 M N + 0.0299488 M^2 N$\\
1,2&$ 0.661318 + 0.410885 M + 0.0401129 M^2+ 0.00432058 MN+ 0.00216029 M^2N$\\
0,3&$0.131463 + 0.0886454\,M + 0.0114569\,M^2$\\
4,0&$-1.157005 - 0.397981\,M\,N - 0.0273887\,M^2 N^2-0.0000135411\,M^3\,N^3$\\
3,1&$  -2.6265 - 1.31325\,M - 0.26635\,M\,N - 0.133175\,M^2\,N + 
   0.00011597\,M^2\,N^2 + 0.0000579851\,M^3\,N^2$\\
2,2&$ -2.31478 - 1.56359\,M - 0.203103\,M^2 - 0.0493674\,M\,N - 
   0.0244594\,M^2\,N + 0.000112162\,M^3\,N$\\
1,3&$ -0.970048 - 0.704633\,M - 0.10947\,M^2 + 0.00016724\,M^3 - 
   0.00381205\,M\,N - 0.00238253\,M^2\,N - 0.000238253\,M^3\,N$\\
0,4&$-0.166668 - 0.124257\,M - 0.0206675\,M^2 - 0.000103042\,M^3$\\
5,0&$2.26883 + 0.901262\,M\,N + 0.0889695\,M^2\,N^2 + 0.00136005\,M^3\,N^3-6.90887 10^{-8} M^4 N^4$\\
4,1&$  6.58636 + 3.29318\,M + 0.970818\,M\,N + 0.485409\,M^2\,N + 
   0.0167572\,M^2\,N^2 + 0.00837862\,M^3\,N^2 $\\&$+ 0.000014758\,M^3\,N^3 + 
   7.37899\,{10}^{-6}\,M^4\,N^3$\\
3,2&$ 8.02544 + 5.66644\,M + 0.826859\,M^2 + 0.359772\,M\,N+0.221805\,M^2\,N+ 
   0.0209597\,M^3\,N + 0.00113766\,M^2\,N^2 $\\&$+ 0.000617202\,M^3\,N^2 + 
   0.0000241853\,M^4\,N^2$\\
2,3&$ 5.23732 + 4.0716M+ 0.770463M^2 + 0.0219952M^3 + 0.058081MN + 
   0.039797M^2N + 0.0054257M^3N + 0.000023694 M^4 N$\\
1,4&$1.84567 + 1.47827\,M + 0.302198\,M^2 + 0.0122745\,M^3+0.000017524\,M^4 + 
   0.00491992\,M\,N + 0.00347\,M^2\,N $\\&$+ 0.000475809\,M^3\,N - 
   0.0000146041\,M^4\,N$\\
0,5&$  0.278724 + 0.226356\,M + 0.0478936\,M^2 + 0.00219097\,M^3 - 
   3.68575\,{10}^{-6}\,M^4$\\
\end{tabular}
\end{table}

\begin{table}[tbp]
\squeezetable
\caption{The coefficients $b^{(v)}_{ij}$, cf. Eq. (\ref{bv}).
}
\label{t2}
\begin{tabular}{cl}
\multicolumn{1}{c}{$i,j$}&
\multicolumn{1}{c}{$R_{NM}^{-i} R_M^{-j} b^{(v)}_{ij}$}\\
\tableline \hline
%beta v
2,0&$-1.14424 - 0.0720268\,M\,N$\\
1,1&$-1.62169 - 0.310751\,M$\\
0,2&$-0.588581 - 0.127593\,M$\\
3,0&$ 1.68536 + 0.162556\,M\,N - 0.00251242\,M^2\,N^2$\\
2,1&$ 3.65454 + 0.776108\,M + 0.036274\,M\,N - 0.0011187\,M^2\,N$\\
1,2&$2.74577 + 0.729746\,M + 0.0159795\,M^2$\\
0,3&$0.719309 + 0.204598\,M + 0.00685909\,M^2$\\
4,0&$ -3.15852 - 0.410304\,M\,N-0.00388438\,M^2\,N^2 - 0.00012569\,M^3\,N^3$\\
3,1&$ -9.25529 - 2.09353\,M - 0.273324\,M\,N - 0.0473612\,M^2N + 
   0.00103599\,M^2\,N^2 - 0.000247572 M^3 N^2$\\
2,2&$-10.5999 - 3.25027\,M - 0.159554\,M^2 - 0.0210851\,M\,N - 
   0.00248042\,M^2\,N - 0.0000643977\,M^3\,N$\\
1,3&$-5.61836 - 1.86565\,M - 0.116276\,M^2 + 0.0000353371\,M^3$\\
0,4&$-1.157 - 0.397981\,M - 0.0273887\,M^2 - 0.0000135411\,M^3$\\
5,0&$ 7.02662 + 1.14357\,M\,N + 0.0316503\,M^2\,N^2 - 0.000233057\,M^3\,N^3 - 
   7.79352\,{10}^{-6}\,M^4\,N^4$\\
4,1&$  26.007 + 6.15895\,M + 1.33378\,M\,N + 0.286892\,M^2\,N - 
   0.0021517\,M^2\,N^2 - 0.00120819\,M^3\,N^2 $\\&$+ 0.0000320327\,M^3\,N^3 - 
   0.0000252216\,M^4\,N^3$\\
3,2&$  40.1395 + 13.492\,M + 0.87123\,M^2 +0.403601\,M\,N+0.0984158\,M^2\,N - 
   0.00021126\,M^3\,N - 0.000514422\,M^2\,N^2 $\\&$- 
   8.56408\,{10}^{-6}\,M^3\,N^2 - 0.0000250754\,M^4\,N^2$\\
2,3&$32.1867 + 12.0427 M + 1.0329 M^2 +0.00813502\,M^3 - 0.00699968 M N - 
   0.004025 M^2 N $\\&$- 0.000484565 M^3N - 3.956\,{10}^{-6} M^4 N$\\
1,4&$ 13.3343 + 5.18122\,M + 0.485923\,M^2 + 0.00596934\,M^3 + 
   3.27121\,{10}^{-6}\,M^4$\\
0,5&$ 2.26883 + 0.901262\,M + 0.0889695\,M^2 + 0.00136005\,M^3 - 
   6.90887\,{10}^{-8}\,M^4$\\
\end{tabular}
\end{table}

%eta

\begin{table}[tbp]
\squeezetable
\caption{
The coefficients $e^{(\phi)}_{ij}$, cf. Eq.(\ref{etaphi}).
}
\label{t3}
\begin{tabular}{cl}
\multicolumn{1}{c}{$i,j$}&
\multicolumn{1}{c}{$R_{NM}^{-i} R_M^{-j}e^{(\phi)}_{ij}$}\\
\tableline \hline
2,0&$0.0226441 + 0.011322\,M\,N$\\
1,1&$0.0452882 + 0.0226441\,M$\\
0,2&$0.0226441 + 0.011322\,M$\\
3,0&$-0.00119855 - 0.000749094\,M\,N - 0.0000749094\,M^2\,N^2$\\
2,1&$-0.00359565 - 0.00179783\,M - 0.000449457\,M\,N - 0.000224728\,M^2\,N$\\
1,2&$-0.00359565 - 0.00224728\,M - 0.000224728\,M^2$\\
0,3&$-0.00119855 - 0.000749094\,M - 0.0000749094\,M^2$\\
4,0&$0.00633062 + 0.00445834\,M\,N + 0.000618261\,M^2\,N^2 - 
     0.0000141266\,M^3\,N^3$\\
3,1&$0.0253225 + 0.0126612\,M + 0.00517211\,M\,N + 0.00258606\,M^2\,N - 
     0.000113013\,M^2\,N^2 - 0.0000565063\,M^3\,N^2$\\
2,2&$0.0379837 + 0.0254875\,M + 0.00324782\,M^2 + 0.00126253\,M\,N + 
     0.000461744\,M^2\,N - 0.0000847595\,M^3\,N$\\
1,3&$0.0253225 + 0.0178334\,M + 0.00247304\,M^2 - 0.0000565063\,M^3$\\
0,4&$0.00633062 + 0.00445834\,M + 0.000618261\,M^2 - 0.0000141266\,M^3$\\
5,0&$-0.00722953 - 0.00550947\,M\,N - 0.000978196\,M^2\,N^2 - 
     0.0000178225\,M^3\,N^3 - 1.20103\,{10}^{-6}\,M^4\,N^4$\\
4,1&$-0.0361476 - 0.0180738\,M - 0.00947355\,M\,N - 0.00473677\,M^2\,N - 
     0.000154204\,M^2\,N^2 - 0.0000771022\,M^3\,N^2 $\\&$- 
     0.0000120103\,M^3\,N^3 - 6.00517\,{10}^{-6}\,M^4\,N^3$\\
3,2&$-0.0722953 - 0.0502342\,M - 0.0070433\,M^2 - 0.00486051\,M\,N - 
     0.00273867\,M^2\,N - 0.000154207\,M^3\,N + $\\&$
     5.95387\,{10}^{-9}\,M^2\,N^2 - 0.0000240177\,M^3\,N^2 - 
     0.0000120103\,M^4\,N^2$\\
2,3&$-0.0722953 - 0.0544783\,M - 0.00942071\,M^2 - 0.000127693\,M^3 - 
     0.000616447\,M\,N - 0.000361246\,M^2\,N $\\&$- 0.0000505316\,M^3\,N - 
     0.0000120103\,M^4\,N$\\
1,4&$-0.0361476 - 0.0275474\,M - 0.00489098\,M^2 - 0.0000891125\,M^3 - 
     6.00517\,{10}^{-6}\,M^4$\\
0,5&$-0.00722953 - 0.00550947\,M - 0.000978196\,M^2 - 0.0000178225\,M^3 - 
     1.20103\,{10}^{-6}\,M^4$\\
\end{tabular}
\end{table}

\begin{table}[tbp]
\squeezetable
\caption{
The coefficients $e^{(t)}_{ij}$, cf. Eq.(\ref{etat}).
}
\label{t4}
\begin{tabular}{cl}
\multicolumn{1}{c}{$i,j$}&
\multicolumn{1}{c}{$R_{NM}^{-i} R_M^{-j} e^{(t)}_{ij}$}\\
\tableline \hline
%eta2
2,0&$0.166698 + 0.0833489\,M\,N$\\
1,1&$0.333395 + 0.166698\,M$\\
0,2&$0.166698 + 0.0833489\,M$\\
3,0&$-0.132662 - 0.0893945\,M\,N - 0.0115318\,M^2\,N^2$\\
2,1&$-0.397986 - 0.198993\,M - 0.0691908\,M\,N - 0.0345954\,M^2\,N$\\
1,2&$-0.397986 - 0.259542\,M - 0.0302748\,M^2 - 0.00864116\,M\,N - 
     0.00432058\,M^2\,N$\\
0,3&$-0.132662 - 0.0893945\,M - 0.0115318\,M^2$\\
4,0&$0.172999 + 0.128715\,M\,N + 0.0212857\,M^2\,N^2 + 0.0000889159\,M^3\,N^3$\\
3,1&$0.691995 + 0.345998\,M + 0.168863\,M\,N + 0.0844315\,M^2\,N + 
     0.000711327\,M^2\,N^2 + 0.000355664\,M^3\,N^2$\\
2,2&$1.03799 + 0.711189\,M + 0.0960966\,M^2 + 0.0611014\,M\,N + 
     0.0316177\,M^2\,N + 0.000533495\,M^3\,N$\\
1,3&$0.691995 + 0.507236\,M + 0.0803778\,M^2 - 0.000120843\,M^3 + 
     0.0076241\,M\,N + 0.00476506\,M^2\,N + 0.000476506\,M^3\,N$\\
0,4&$0.172999 + 0.128715\,M + 0.0212857\,M^2 + 0.0000889159\,M^3$\\
5,0&$-0.285954 - 0.231865\,M\,N - 0.0488718\,M^2\,N^2 - 0.0022088\,M^3\,N^3 + 
     2.48471\,{10}^{-6}\,M^4\,N^4 $\\ 
4,1&$-1.42977 - 0.714885\,M - 0.444443\,M\,N - 0.222221\,M^2\,N - 
     0.0221377\,M^2\,N^2 - 0.0110688\,M^3\,N^2 $\\&$+ 0.0000248471\,M^3\,N^3 + 
     0.0000124236\,M^4\,N^3$\\
3,2&$-2.85954 - 2.02478\,M - 0.297507\,M^2 - 0.293871\,M\,N - 
     0.189024\,M^2\,N - 0.0210442\,M^3\,N $\\&$- 0.00218691\,M^2\,N^2 - 
     0.00104376\,M^3\,N^2 + 0.0000248471\,M^4\,N^2$\\
2,3&$-2.85954 - 2.23824\,M - 0.433491\,M^2 - 0.0146269\,M^3 - 
     0.0804108\,M\,N - 0.055227\,M^2\,N $\\&$- 0.0074611\,M^3\,N + 
     0.0000248471\,M^4\,N$\\
1,4&$-1.42977 - 1.14949\,M - 0.237419\,M^2 - 0.0100924\,M^3 - 
     0.0000167847\,M^4 - 0.00983986\,M\,N - 0.00694\,M^2\,N $\\&$- 
     0.00095162\,M^3\,N + 0.0000292083\,M^4\,N$\\
0,5&$-0.285954 - 0.231865\,M - 0.0488718\,M^2 - 0.0022088\,M^3 + 
     2.48471\,{10}^{-6}\,M^4 $\\
\end{tabular}
\end{table}

\subsection{Resummations of the series and analysis method}
\label{sec2c}

The obtained RG series are asymptotic and some resummation procedure 
is needed in order to extract accurate numerical values for the 
physical quantities. Exploiting the property of Borel summability of 
$\phi^4$ theories in two and three dimensions \cite{Borsum}, 
we resum the divergent 
asymptotic series by a Borel transformation combined with an
analytic extension of the Borel transform to the real positive axis. 
This extension can be obtained by a Pad\'e approximant or by a conformal 
mapping \cite{LZ-77} which maps the domain of analyticity of the 
Borel transform onto a circle 
(see Refs.~\cite{LZ-77,ZJ-book} for details).

The conformal mapping method takes advantage of the knowledge of the 
large order behavior of the perturbative series 
$F(\ub,z)=\sum_k f_{k}(z) \ub^k $~\cite{LZ-77,ZJ-book}
\begin{equation}
f_k(z) \sim k! \,(-a(z))^{k}\, k^b \,\left[ 1 + O(k^{-1})\right] 
\qquad
{\rm with}\qquad a(z) = - 1/\ub_b(z),
\label{lobh}
\end{equation}
where $\ub_b(z)$ is the singularity of the Borel transform closest to the 
origin at fixed $z= \bar{v}/\bar{u}$, given by \cite{CPV,CC01,pv-00}
\begin{eqnarray}
{1\over \ub_b (z)} &= - a \left( R_{M N}  + R_M z \right)
\qquad & {\rm for} \qquad  0< z \quad {\rm and } 
\quad z<-{2 N \over N+1}\frac{R_{MN}}{R_M},
\label{bsing} \\
{1\over \ub_b (z)} &= - a \left( R_{M N}  + {1\over N} R_M z \right)
\qquad & {\rm for} \qquad  0 > z > - {2 N \over N+1}\frac{R_{MN}}{R_M},
\nonumber
\end{eqnarray}
where $a = 0.238659217\dots \,$ \cite{ZJ-book}.
Note that the condition of Borel summability (that coincides with the 
mean-field boundness condition) is $z>z_1=-R_{MN}/R_M$ for $\vb<0$. 
So in this region, even if the second of Eq. (\ref{bsing}) takes into account
the singularity of the Borel transform closest to the origin, there is 
another singularity on the real positive axis that makes the series not Borel 
summable. The situation is the same as for $O(M)\times O(N)$ under 
$v_{\rm ch}>0$ \cite{chiral3d,cp-01,cps-02}. 
As in the latter case we resum the series even where they are 
not Borel summable. Although in this case the sequence of approximation is 
only asymptotic, it should provide reasonable estimates as long as we are 
taking into account the leading large-order behavior (i.e. as long as 
$z > z_2=- {2 N \over N+1}\frac{R_{MN}}{R_M}$).

These results do not apply to the case $N=0$. Indeed, in this case, 
additional singularities in the Borel transform are expected \cite{nonbs}.

For each perturbative series $R(\bar{u},\bar{v})$,
we consider the following approximants \cite{LZ-77}
\begin{equation}
E({R})_p(\alpha,b;\ub,\vb)= \sum_{k=0}^p 
  B_k(\alpha,b;\vb/ \ub)    \int_0^\infty dt\,t^b e^{-t} 
  \frac{[y(\ub t;\vb/\ub)]^k}{[1 - y(\ub t;\vb/ \ub)]^\alpha},
\label{approx}
\end{equation}
where 
\begin{equation}
y(x;z) = {\sqrt{1 - x/\overline{u}_b(z)} - 1\over
 \sqrt{1 - x/\overline{u}_b(z)} + 1}.
\end{equation}
The coefficients $B_k$ are determined by the condition that the 
expansion of $E({R})_p(\alpha,b;\ub,\vb)$ in powers of $\ub$ and 
$\vb$ gives $R(\ub,\vb)$ to order $p$.
Within this method we end up with several values for each resummed quantity,
depending upon the free parameters $\alpha$ and $b$. 
To have a single estimate we search for the values of $\alpha_{\rm opt}$
and $b_{\rm opt}$ minimizing the difference between the highest orders
and, as usual \cite{LZ-77,CPV,CC01}, we consider as final estimate 
(error bar) the average (the variance) of the approximants with 
$\alpha\in [\alpha_{\rm opt}-\Delta\alpha,\alpha_{\rm opt}+\Delta\alpha]$
and $b\in [b_{\rm opt}-\Delta b,b_{\rm opt}+\Delta b]$, with
$\Delta\alpha=2$ and $\Delta b=3$ (see e.g. Ref. \cite{CPV} for a discussion
about the effectiveness of such choice of $\Delta\alpha$ and $\Delta b$).

Within the second resummation procedure, the Borel-Leroy transform is 
analytically extended by means of a generalized Pad\'e approximant, 
using the resolvent series trick (see, e. g. \cite{PS}).
Explicitly, once 
introduced the resolvent series of the perturbative 
one $R(\bar{u},\bar{v})$
\begin{equation}
\tilde{P}(R)(\ub,\vb,b,\lambda)=\sum_{n} \lambda^n\sum_{k=0}^n 
{\tilde{P}_{k,n} \over \Gamma(n+b+1)} \ub^{n-k}\vb^{k}\, , 
\end{equation}
which is a series in powers of $\lambda$ with coefficients being 
uniform polynomials in $\ub,\vb$. The analytical continuation of the 
Borel transform is the Pad\`e approximant $[N/M]$ in $\lambda$ at 
$\lambda=1$. Obviously, the sum for each perturbative series 
depends on the chosen Pad\'e approximant and on the parameter $b$.
We consider for the final estimates all the non-defective (i.e. having no  
singularities on the real positive axis) Pad\'e approximants at the value 
of $b$ minimizing the difference between them.

An important issue in the fixed dimension approach to critical 
phenomena~(and in general of all the field theoretical methods) 
concerns the analytic properties of the RG functions. 
As shown in Ref. \cite{CCCPV} for the $O(N)$ model, the presence of 
confluent singularities in the zero of the perturbative $\beta$ function
causes a slow convergence of the results given by the resummation of the 
perturbative series to the correct fixed point value. The $O(N)$ 
two-dimensional field-theory estimates of physical quantities \cite{LZ-77,OS}
are less accurate than the three-dimensional ones \cite{LZ-77} 
due to the stronger non-analyticities at the fixed point
\cite{CCCPV,Nick91,PV98}, to say nothing about the stronger growth of 
the series coefficients themselves.
In Ref. \cite{CCCPV} it is shown that the non-analytic terms may cause
large imprecision in the estimate of the exponent related to the leading 
correction to the scaling $\omega$. At the same time, the result for the 
fixed point location turns out to be a rather good approximation 
compared with those coming from different techniques.
Non-analyticities cause slow convergence to the exact critical exponents too.

\section{ANALYSIS OF Five Loop series}
\label{sec3}

\subsection{Stability of the $O(NM)$ and $O(M)$ fixed points}

We start the analysis of five-loop perturbative series in the case when one
coordinate of the fixed point (FP) vanishes. Since in this case a lot 
of exact results are known, these calculations will be a check of the goodness 
of our resummation procedure and of the error made considering {\it only} 
five loops in the perturbative expansion.

The location of the FP on the axis $\vb=0$ (that is equivalent to $\ub=0$, 
with replacing $NM\rightarrow M$) was already studied in Ref. \cite{OS}. 
Also the critical exponents were considered there. 
The agreement of these results with the available 
exact ones was satisfactory (see also footnote \cite{OS-corr}). 
What still remains to be computed is the 
stability of the FP's with respect to perpendicular perturbation.

First of all, we  analyze the stability properties of the $O(N M)$-symmetric
fixed point. 
Since $ \partial_\ub \beta_\vb |_{(\ub,0)} = 0$, the stability 
with respect to an anisotropic cubic perturbation is given by 
$\omega_v(\ub)=2\partial_\vb\bar{\beta}_\vb|_{(\ub,0)}$,
whose five-loop expansion is ($n=NM$)
\begin{eqnarray}
\frac{\omega_v(\ub)}{2}&=& 
-1+\frac{12}{n+8} \ub-\frac{92.6834 + 5.83417 n}{(n+8)^2}\ub^2 
+\frac{1228.63 + 118.503\,n - 1.83156\,n^2}{(n+8)^3}\ub^3\nonumber\\&&
-\frac{20723.1 + 2692.00 n + 25.4854 n^2 - 0.824655 n^3}{(n+8)^4}\ub^4\nonumber\\&&
+\frac{414915.7 + 67526.8\,n + 1868.92\,n^2 - 13.7618\,n^3 - 0.4602\,n^4}{(n+8)^5}\ub^5\,.
\label{omv}
\end{eqnarray}
This series must be calculated at the $O(n)$ FP located at $\ub=\ub^*$.
In Table \ref{omon} we report the results for $\omega_v$ for several values of
$n$, obtained resumming the series (\ref{omv}) at the FP's calculated 
in Ref. \cite{OS}.
The $O(n)$ FP results unstable for $n \geq 3$.
For $n=2$ our result is compatible with the expected result $\omega_v=0$, 
which is essential in the context of a continuous line of the fixed points.

We can also use these results to discuss the nature 
of the multicritical point in two-dimensional models with symmetry 
O$(N_1)\oplus {\rm O}(N_2)$ \cite{KNF}.  
Indeed, they allow to  exclude that the multicritical point has enlarged 
symmetry $O(N_1 + N_2)$ if $N_1 + N_2 =n> 2$ \cite{cpvmult}. 
In RG terms, this can generally occur if
the $O(N_1 + N_2)$ fixed point has only two relevant
O($N_1$)$\oplus$ O($N_2$)-symmetric perturbations. 
But, when $n>2$, the instability of the O($n$) fixed point with respect to 
the cubic perturbation shows that at least one extra 
relevant perturbation exists (see for details Ref. \cite{cpvmult}).

Then we focus our attention on  the stability properties of the $O(M)$
fixed point. 
Also in this case the stability is given by 
$\omega_u(\vb)=2\partial_\ub\bar{\beta}_\ub|_{(0,\vb)}$,
evaluated at the $O(M)$ fixed point $\vb^*$.
As expected, the series $\omega_u(\vb)$ is independent of $N$
\begin{eqnarray}
{\omega_u(\vb)\over 2}&=&-1+2\frac{M+2}{M+8} \vb -
\frac{5.83417 (M+2)}{(M+8)^2}\vb^2
+ \frac{(47.9183 + 8.35207 M)(M+2)}{(M+8)^3} \vb^3\nonumber\\&&
-\frac{(546.755 + 134.247 M + 0.67606 M^2)(M+2)}{(M+8)^4}\vb^4\nonumber\\&&
+\frac{(8229.19 + 2568.45 M + 129.81 M^2 - 0.2176 M^3)(M+2)}{(M+8)^5}\vb^5.
\label{omu}
\end{eqnarray}
For $M=1$, the fixed point value of this exponent is $\omega_u^I/2=-0.09(8)$,
using the conformal mapping method, and $-0.08(10)$ with the Pad\'e-Borel. 
We note that the [4/1] approximant with $b=1$ leads to $\omega_u^I/2=-0.031$.
These values are compatible with the exact known result $\alpha_I=0$.
For $M\geq2$, we always find $\omega_u>0$, in agreement with the exact 
results predicting the stability of the $O(M)$ fixed point. 

\begin{table}[t]
\caption{Half of the exponent $\omega_v$ at the $O(n)$ fixed point. 
CM is the value 
obtained using conformal mapping technique and PB the one using Pad\'e-Borel.}
\begin{tabular}{r|cc|cc}
n &  CM 4-loop &  PB 4-loop &  CM5-loop & PB 5-loop\\
\tableline \hline
2  &$ 0.03(3)$  & $ 0.06(4)$ & $0.025(40)$ & $0.00(5)$ \\
3  &$-0.08(3)$  & $-0.07(3)$ & $-0.10(6) $ & $-0.10(5)$ \\
4  &$-0.18(4)$  & $-0.17(5)$ & $-0.17(4) $ & $-0.20(4)$ \\
8  &$-0.45(5)$  & $-0.44(6)$ & $-0.48(4) $ & $-0.50(5)$  
\end{tabular}
\label{omon}
\end{table}

%%%%%%%%%%%%%%%%%%

An alternative approach to the analysis of the series Eqs (\ref{omv}) and
(\ref{omu}) is the so called pseudo-$\e$ expansion \cite{pseudo}. 
Within this method, one multiplies the linear term in the 
$\beta$ functions by a fictitious small
parameter $\tau$, finds the common zeros of the $\beta$ functions as 
series in $\tau$, and finally the exponents (as series in $\tau$) are 
obtained introducing such expansions in the corresponding RG functions. 
This approach has twofold advantage. First, in the estimates of the critical 
exponents the cumulation of errors due to not exact knowledge of the FP
and that of the proper uncertainty of the RG function is avoided 
\cite{LZ-77}.
Second, it allows for very precise estimates of the marginal number of 
components since these are series in $\tau$ that must be evaluated at 
$\tau=1$ \cite{HDY-01,dhy-04,cp-03,cp-04}. 
We use this method only for the second purpose.

Imposing the vanishing of $\omega_v$ at the $O(NM)$ fixed point, the 
series for $N_c$ determining the relevance of cubic 
anisotropy (see the Introduction) is:
\begin{equation}
N_c=4 - 2.2504 \tau + 0.6230 \tau^2 - 0.7509 \tau^3 + 1.1761 \tau^4.
\label{Nc}
\end{equation}
Analogously, imposing $\omega_u=0$ at the $O(M)$ FP leads to
\begin{equation}
M_c=4 - 4.5008 \tau + 2.1693 \tau^2 - 1.2165 \tau^3 + 1.3055 \tau^4,
\label{Mc}
\end{equation}
determining the relevance of an energy-like interaction, as e.g. the relevance
of quenched randomness.
Such series do not behave asymptotically with a factorial growth of the
coefficients (at least up to the considered order), 
thus simple Pad\'e approximants without Borel resummation are expected to
give reliable estimates. This in fact turned out to be the case for their
three dimensional analogs \cite{HDY-01,dhy-04} and also for 
more complicated series as those for the marginal number of components 
of $O(N)\times O(M)$ \cite{cp-03} and $U(N)\times U(M)$ \cite{cp-04} models.

The Pad\'e table for the series Eq. (\ref{Nc}) at $\tau=1$ is
\begin{equation}
\left[
\begin{array}{ccccc}
4    &1.75 & 2.37 & 1.62& 2.53\\
2.56 &2.24 & 2.03 & 2.08\\
2.32 &5.76_{[0.9]}& 2.08\\
2.08 &2.17\\
2.21_{[3.1]}\\
\end{array}
\right]\,,
\end{equation}
where we indicated in small brackets the closest pole to the origin on the 
real axis (when exists). 
Whenever this pole is close to $\tau=1$, the approximant has not to be
considered in the average procedure. This Pad\'e table leads to 
the estimates $N_c=2.10(7)$ that includes all the four- and five-loop 
estimates without poles (excluding the first line and column that 
are unreliable). This is in quite good agreement with the known exact 
result $N_c=2$ \cite{pn-76}, signaling about the high effectiveness of 
the perturbative expansion technique at the five-loop level.

Similarly, the Pad\'e table for the series Eq. (\ref{Mc}) is
\begin{equation}
\left[
\begin{array}{ccccc}
4    & -0.5 & 1.66 & 0.45 & 1.75\\
1.88 & 0.96 & 0.89 & 1.08\\
1.40 & 0.88_{[8.2]} & 0.95\\
1.19 & 1.09\\
1.13 \\
\end{array}
\right]\,.
\end{equation}
The same averaging procedure as before leads to $M_c=0.99(10)$, that 
perfectly agrees with the exact value $M_c=1$ at which the specific-heat 
exponent is vanishing. This demonstrates under $\ub=0$ good approximating 
properties of the five-loop series.

We analyzed the series (\ref{Nc}) and (\ref{Mc}) with the Pad\'e-Borel
resummation as well, having obtained equivalent results. So, indeed at 
the considered order these series do not behave as asymptotic.

\subsection{The cubic model for $N\geq 3$}
\label{sec3b}

In this Section we consider the existence of the cubic fixed point for 
$N\geq 3$, previously found in the four-loop approximation \cite{CC01}.
It was claimed in Ref. \cite{CC01} that the quite peculiar features of
this fixed point (like the marginal instability) make its existence quite 
doubtful and that it might be an artifact of the {\em relatively} 
short series available at that time. Now we are in a position to analyze
longer series and to further confirm or to reject this statement.

The results obtained with the conformal mapping
methods are reported in Table \ref{tabn>3} together with the previous 
four loop results \cite{CC01}, 
in order to make the comparison visible at first sight.
This Table shows that the cubic fixed point drastically moves towards 
the Ising fixed point with increasing the order of perturbation 
theory from four to five loops. It may be considered as an argument 
in favor of the statement that, within the exact theory, the cubic 
and the Ising fixed points coincide. 
On the other hand, both at four and at five loops, the quoted errors are
less than the difference between the two estimates. This leads to the conclusion
that the reported uncertainty is actually an underestimate of the correct one.
We remind to the reader that this error come from the so called 
stability criterion, i. e. it is obtained looking at those 
approximants that minimize the difference between the estimates at the
highest available orders (see Sec. \ref{sec2c}). So, the considerable 
discrepancies between the four- and five-loop results lead to serious doubts 
in the existence of the cubic fixed point in two dimensions.

\begin{table}[t]
\caption{Location of the apparent cubic fixed point for some $N\geq3$.}
\begin{tabular}{r|ccc}
$N$ & CM 4-loop & CM 5-loop &  Pad\'e [4/1] $b=1$ \\
\tableline \hline
3 & [0.83(12),1.12(9)] & [0.54(6),1.35(4)]  & [0.050,1.757] \\
4 & [0.54(10),1.43(8)] & [0.32(5),1.58(4)]  & [0.031,1.774] \\
8 & [0.24(8),1.72(10)] & [0.14(4),1.74(4)]  & [0.015,1.788]   
\end{tabular}
\label{tabn>3}
\end{table}

\begin{figure}[t]
%\vspace{-1.5cm}
\centerline{\epsfig{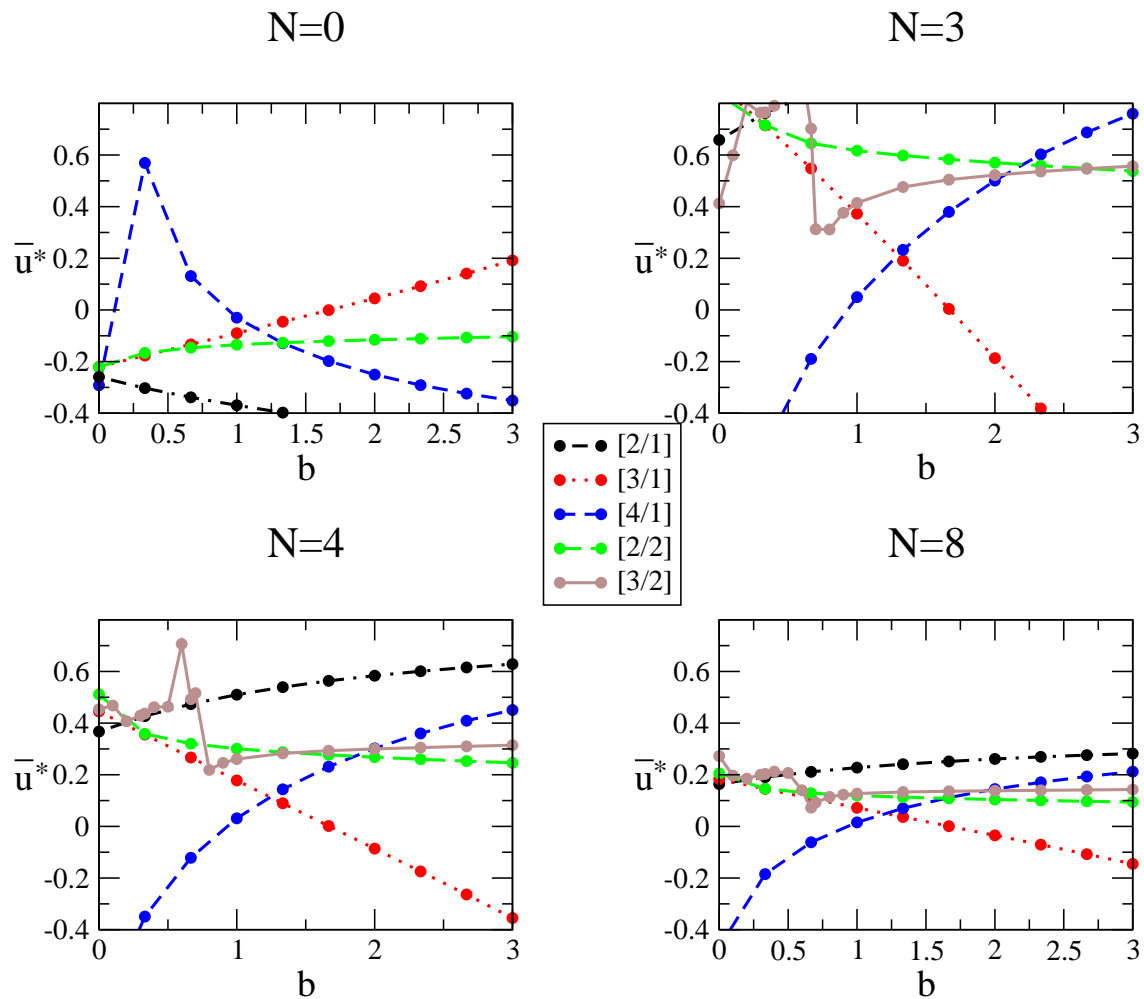}}
%\vspace{-1.5cm}
\caption{Coordinate $u^*$ of the cubic fixed point within Pad\'e-Borel method
for some values of N.}
\label{padefig}
\end{figure}

In order to understand better these somewhat strange results, we report
the values of the coordinate $\ub^*$ obtained for the cubic fixed point 
using several Pad\'e approximants for $N = 3, 4, 8$; the estimates  
are presented in Fig. \ref{padefig} as functions of $b$.
Let us consider first the case $N=3$ as a typical example. If one 
limits himself with only three lower-order approximants [2/1], [3/1], 
and [2/2], he easily finds that they minimize their differences under 
$b\sim 0$, leading in such a way to the estimate $\ub^* \sim 0.7$. 
Taking into account two non-defective Pad\'e approximants [4/1] and 
[3/2] (note the oscillating behavior of the [3/2] approximant for $b<1$, 
signaling the presence of close singularities) existing at the five-loop 
level shifts the zone of stability  
to $b\sim 2$, thus leading to the estimate $\ub^*\sim 0.5$. 
Moreover, the approximant [4/1] with $b=1$, that is usually  
considered as one of the best approximants, results in the estimate 
$\ub^*=0.050$ very close to zero. Because of the alternative character 
of RG expansions, it looks very likely that the unknown 
six-loop contribution (and the higher-order ones) will locate the 
stability region somewhere near $b\sim 1$, leading finally to the
coalescence of the Ising and cubic fixed point. According to this 
scenario, the cubic fixed point, found at finite order in perturbation 
theory, is probably only an artifact due to the finiteness of the perturbative
series.

The same scenario is possible also for other values of $N$. From 
Fig. \ref{padefig} we see that the region of maximum stability always
shifts from $b\sim 0$ to $b\sim 2$ with increasing the order of approximation 
from four to five loops, moving the coordinate of the fixed point from 
$\ub^*\sim0.5$ for $N=4$ ($\ub^*\sim0.2$ for $N=8$) to $\ub^*\sim0.3$ 
($\ub^*\sim0.1$). Let us stress again that the value given by the approximant 
[4/1] with $b=1$ is always very close to zero, as is seen from 
Table \ref{tabn>3}. Note that, with increasing $N$, the distance between 
the cubic fixed point and the Ising one reduces rapidly.

In the limit $N\rightarrow\infty$ the series simplify as at the four-loop 
level (see Ref. \cite{CC01}). We only mention that with increasing the length 
of the RG series the coordinate $\ub^*$ of the cubic fixed point shifts from 
$\ub^*\sim 0.08$ to $\ub^* \sim0.03$ that again is much more close to zero.

\subsection{The random $O(M)$ model}

The $MN$ model in the limit $N\rightarrow0$ describes the critical behavior 
of quenched dilute $O(M)$ models. Being $u_0$ proportional to minus the 
variance of the disorder \cite{rev-01,revran} 
only the region with $\ub<0$ is of physical interest.
We have already shown that for $M\geq2$ the $O(M)$ fixed point is stable.
To analyze the RG flow in the whole physical region we do not use here 
advanced resummation procedures developed \cite{pv-00} to avoid Borel 
non-summability at fixed $u/v$\cite{nonbs}, but limit ourselves by a simple 
Pad\'e analysis, since it is sufficient for our aims.
Within this method we check for several $M\geq2$ that no other FP with 
$\ub<0$ exists when the resummation is effective. Thus, whenever the 
transition of dilute models is second-order (i.e. under the percolation 
threshold), the critical behavior is of $O(M)$ type.
An unstable FP is found in the unphysical region $\ub,\vb>0$, which delimits
the domain of attraction of the $O(M)$ and $O(0)$ fixed points, in agreement 
with the extrapolation of $\e$-expansion at $\e=2$ \cite{Aharony-76}.

Finally, we discuss the fate of the random fixed point governing 
the critical behavior of the weakly-disordered Ising model ($M=1$ and $N=0$)
that is found in three dimensions \cite{PS,pv-00,revran} and close to four 
dimensions by means of the $\sqrt\e$ expansion \cite{Kh-75,SAS,revran}.
In this case the Ising FP is marginally stable for $\ub>0$ \cite{Sh94}, 
so the RG flow is driven to $\ub^*=0$ from the physical region (in the 
unphysical region the Ising FP is marginally unstable and the RG flows 
run into the $O(0)$ FP).
We find, for the majority of the considered approximants, a fixed point with 
negative $\ub^*$, as reported in Fig. \ref{padefig}. 
A possible estimate, according to stability criteria is 
$\ub^*=-0.1(1)$, but if we concentrate on some certain approximants we obtain 
$\ub^*=-0.090$ for the [3/1] and $\ub^*=-0.030$ for the [4/1] (both 
with $b=1$). 
In particular, the last value is very close to zero, i. e. to the value 
predicted by the asymptotically exact solution that has been obtained in the 
framework of the fermionic representation \cite{DD83,Sh84,Sh94}. 
It is worthy to note that the five-loop results for $N=0$ seem to be less 
scattered than analogous four-loop estimates obtained by means of  
Chisholm-Borel resummation technique \cite{MSS}, and they look more 
precise than their five-loop counterparts for finite $N$.

\subsection{The cubic model for $N=2$}
\label{sec3c}

The four-loop analysis of Ref. \cite{CC01} for $N=2$ turned out to be 
compatible with the presence of a line of the fixed points joining the 
$O(2)$-symmetric and the decoupled Ising fixed points.
The lines of zeros of the two $\beta$ functions were found to be 
practically parallel and the quoted error was bigger than distance between 
them. This line of fixed points with continuously varying critical 
exponents is in agreement with what is expected from the correspondence, 
at the critical point, between the cubic model and the Ashkin-Teller and 
the planar model with fourth-order anisotropy \cite{JKKN,CC01}. 
We are now in a position to verify this statement at the five-loop level.

\begin{figure}[t]
%\vspace{-1.5cm}
\centerline{\epsfig{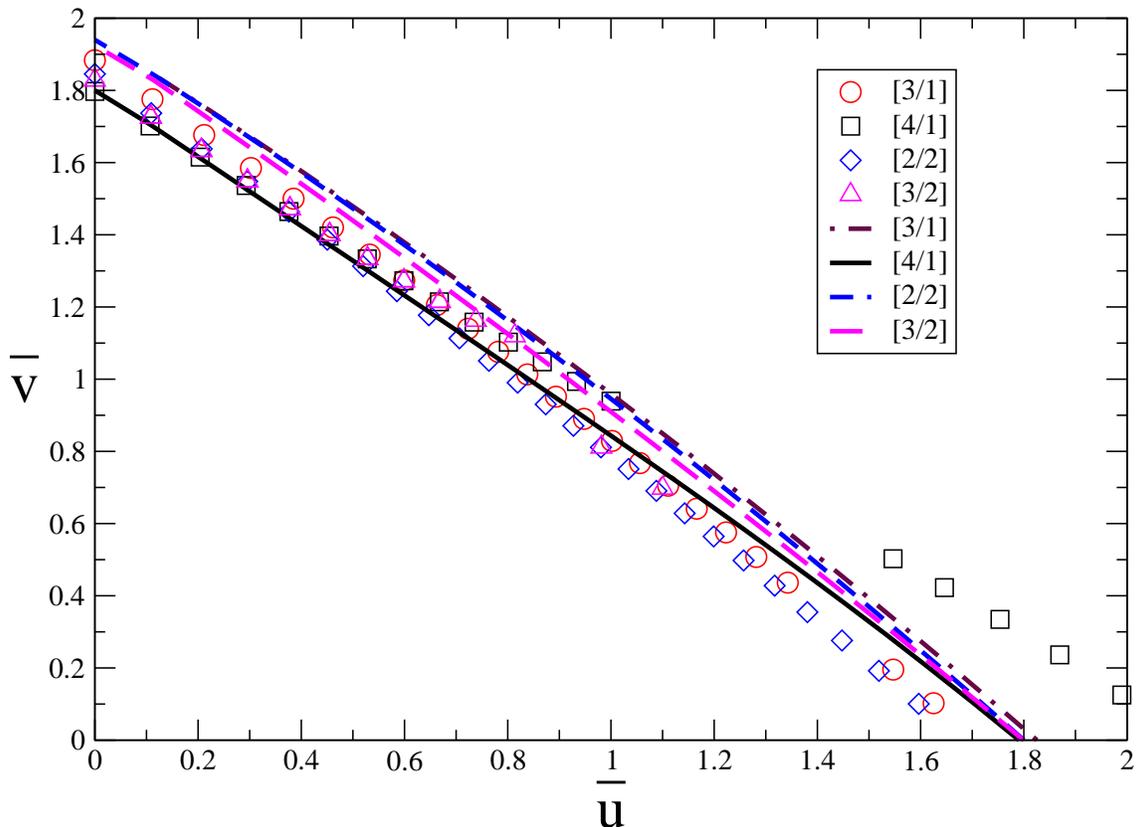}}
%\vspace{-1.5cm}
\caption{Zeros of $\bar{\beta}_\ub$ (continuous lines) and $\bar{\beta}_\vb$ 
(points) for several Pad\'e approximant (all with $b=1$).}
\label{n2}
\end{figure}

First, we analyze the series with the conformal mapping method. Again we find 
that zeros of two $\beta$ functions form two parallel lines, while the 
apparent uncertainties seem to be smaller than their separation. 
Of course, this fact 
may simply indicate that the model has no fixed point at all. 
Let us, however,  
look more carefully to this result and, in particular, to the accuracy of the 
quoted error. In fact, as we have already seen for $N>2$, the error coming 
from stability criteria is likely an underestimate of the correct one. 
To understand better the situation, we use the Pad\'e-Borel method. In 
Fig. \ref{n2} we report the curves of zeros of the two $\beta$ 
functions given by several Pad\'e approximants under $b=1$, the value that 
for $N>2$ always leads to good results and that is the best for the fixed 
point values of $O(N)$ and Ising model \cite{OS}. 
The four approximants for $\bar{\beta}_\ub$ are always well-defined. They are 
hardly distinguishable close to the $O(2)$ fixed point and their separation 
slowly increases moving toward the $\vb$ axis. The coordinate of the Ising 
fixed point $\vb\sim 1.8$ is obtained using the [4/1] approximant, 
since approximants 
of [L-1/1] type proved to give rather precise estimates for the fixed point 
location both in two and three \cite{AS95,S98} dimensions. 
The situation is a bit worse for the function $\bar{\beta}_\vb$.
In fact, the working 
approximants are well defined close to the Ising fixed point, but approaching 
the $\ub$ axis they becomes defective. The approximant [3/2] starts 
oscillating around $\ub\sim0.8$, while [4/1] is bad in the range 
$\ub\in[1,1.5]$ and [3/1] for $\ub > 1.3$. Also the values of zeros of 
the approximant [4/1] for $\ub > 1.5$ are not reliable enough, since they 
may suffer of the effect of close singularities. 

Despite of these shortcomings, we can obtain a rich information from 
Fig. \ref{n2}. Indeed, the line of zeros of $\bar{\beta}_\ub$ given by the 
approximant [4/1] practically coincide with those of the $\bar{\beta}_\vb$ 
from the Ising fixed point up to $\ub\sim 0.8$. For greater $\ub$ various 
approximants for $\bar{\beta}_\vb$ results in the lines of zeros that diverge 
leaving, however, the line [4/1] of $\bar{\beta}_\ub$ zeros between them. 
Keeping in mind a finite length 
of the RG series and the influence of non-analytic terms missed by the 
perturbation theory, we retain this fact as a strong evidence in favor of 
the continuous line of fixed points. The best estimate for this line 
is believed to be that given by the approximant [4/1] for $\bar{\beta}_\ub$. 
Thus it will be used in what follows to calculate continuous varying critical 
exponents.

We evaluate the smallest eigenvalues of the $\Omega$ matrix along the line of
fixed points both with conformal mapping and Pad\'e-Borel method.
We find that it is always compatible with zero (with the uncertainty of the
resummation), that is a necessary condition for having a stable line of fixed
points.

When evaluating the critical exponents, one should keep in mind that 
the limit $z \to 0$ is not simply accessible perturbatively since it 
corresponds to the two-dimensional $XY$ model which is known to behave in a 
quite specific manner. In particular, 
this model does not undergo an ordinary transition into the ordered 
phase at any finite temperature and its critical behavior is essentially 
controlled by the vortex excitations \cite{kt-73}. 
Such excitations lead to an exponentially diverging correlation length at 
finite temperature that can not be accounted for within the 
$\lambda \phi^4$ model Eq. (\ref{Hphi4}) dealt with in this paper.
Since arriving to the point $z=0$ a new physics emerges, it is natural to 
assume that the observables as functions of the fixed point location may be 
non-analytic near this point. Hence, what we can really explore trusting 
upon our five-loop expansions is a domain corresponding to finite 
(and not too small) values of $z$. 
Oppositely, the limit $z \to \infty$ (or $1/z \to 0$) looks quite 
"undangerous" in the above sense since it corresponds to the critical 
behavior close to that of the Ising model, which was shown not to be 
influenced considerably by the non-analytic terms even in two dimensions 
\cite{OS}. Note that just presented results concerning the line of the fixed 
points confirm this idea. Indeed, as seen from Fig. \ref{n2}, at the "Ising 
side", i. e. for $0 < \ub\sim0.8$, the zeros of both $\beta$ functions form 
smooth curves running very close to each other. The closer the "$XY$ side", 
however, the stronger the estimates for $\beta$ function zeros are scattered, 
indicating, likely, the increasing impact of non-analytic contributions.

The expected value of $\eta$ is $1/4$ independently on the location 
of a fixed point within the line. 
The best way to check the constantness of $\eta$ along the line of 
fixed points is probably to write the RG function in terms of $\ub$ and 
$s=\ub+\vb$, i.e. $\eta_s (s,\ub)=\eta (\ub,s-\ub)$. 
Then one resum the difference $\Delta(s,\ub)=\eta_s(s,\ub)-\eta_s(s,0)$. 
Along the line for all the five-loop approximants we always find
$|\Delta(s,\ub)|<8 \times 10^{-3}$. This leads us to conclude that the 
two dimensional LGW approach is able to keep the constantness of $\eta$
within an error of about 3\%.
Note that the previous quoted problems concerning non-analyticities close 
to the $O(2)$ side do not significantly affect the estimates of $\eta$.

\begin{figure}[t]
\centerline{\epsfig{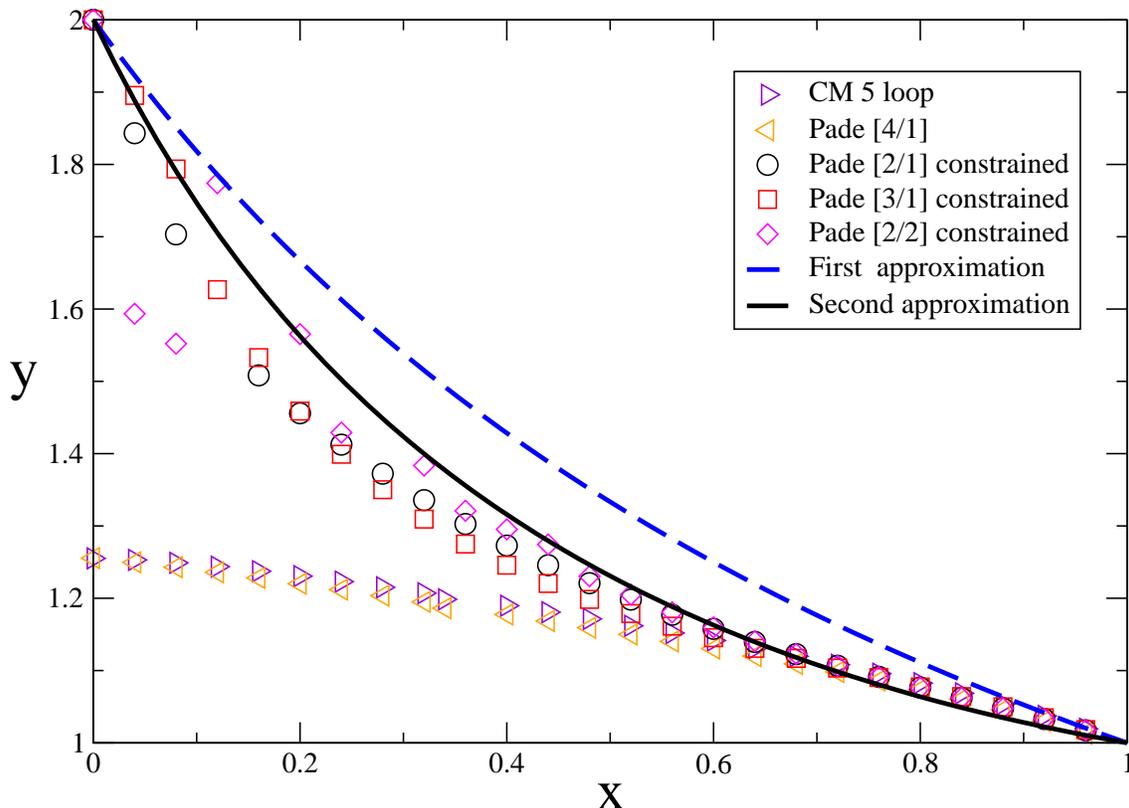}}
\caption{The exponent $y$ as function of $x$ from free and constrained 
resummation. The line "first approximation" is Eq. (\ref{conj}) and the 
line "second approximation" is Eq. (\ref{conj2}).}
\label{yt}
\end{figure}

For the exponent $y=\eta-\eta_t=1/(2-\nu)$ it was conjectured in 
Ref. \cite{CC01} that it should behaves like
\begin{equation}
y={2\over 1+x}, \qquad \mbox{where}, \qquad 
x={2\over \pi} \arctan{\vb^*\over\ub^*}.
\label{conj}
\end{equation}
A direct reliable quantitative estimate of this exponents is impossible
because of the effect of non-analyticities, in particular close to the $O(2)$ 
fixed point for the reasons explained above.
In fact, we know from Ref. \cite{OS,OS-corr} that the resummation of the 
series for $y$ at the Ising fixed point provides $y=1.02$, very close 
to the exact value $1$. Instead, at the $O(2)$ fixed point one has 
$y\sim 1.25$, that is quite far from a diverging $\nu$, i.e. $y=2$.
A direct resummation of this exponent is reported in Fig. \ref{yt}, and as
expected it seems to reproduces the correct critical behavior only
close to the Ising fixed point.

In Ref. \cite{CC01} it was proposed to constrain the exponents to assume the 
exactly known value along the axes to have better quantitative results. 
We apply here a different method of constrained analysis with respect to
Ref. \cite{CC01}. We prefer to constrain the series of $y$ expressed
in terms of $\ub$ and $s=\ub+\vb$ (as previously for the exponent $\eta$),
since the results obtained with this constrain appear more stable.
The constrained Pad\'e-Borel approximants are reported in Fig. \ref{yt}. 
At the Ising side, they are practically indistinguishable from the 
unconstrained ones up to $x\sim 0.6$, but then they start oscillating, 
signaling the presence of singularities leading to bad quantitative estimates.
The conformal mapping results are practically equivalent. 
The numerical data thus obtained are not in agreement with Eq. (\ref{conj}), 
even if this reproduced the right qualitative behavior.
A quadratic form in the denominator of $y$ as e.g. 
\begin{equation}
y={4\over 4-(1-x)(2-x)}\,,  
\label{conj2}
\end{equation}
fits the most of the data much better, as shown in Fig. \ref{yt}.
Unfortunately, we are not able to estimate the goodness of our resummation 
and, as a result, to verify Eq. (\ref{conj2}).
Perhaps, the exact behavior of the exponent $\nu$ along the line of fixed 
points requires new method of analysis of the perturbative series and, in any 
case, it deserves for different studies on the subject like Monte Carlo 
simulation or high temperature expansion.

\subsection{The $MN$ model for $N,M\geq2$}
\label{sec3e}

The critical behavior of the $MN$ model for $M,N\geq2$ and $v_0>0$ can be 
understood by means of non-perturbative arguments. In fact, as we explained
in the Introduction, it turns out that the $O(M)$ fixed point is always stable
and oppositely the $O(MN)$ one is always unstable. Thus the RG flow is 
driven to the $O(M)$ fixed point from both positive and negative $\ub$. The 
only extra fixed point is provided by the $\e$ expansion with $\ub^*<0$ 
which limits the attraction domain of the $O(M)$ one. 
These features are reproduced by
the resummation of the five-loop series. In particular, we found only one
fixed point with non-vanishing coordinates that is the mixed one of the 
$\e$ expansion, that in fact turns out to be unstable.

More interesting is the RG flow for $v_0<0$. In this case the $\e$ expansion 
does not provide any fixed points, thus the transition (if any) cannot be
continuous. Nevertheless for $M=N=2$ we can relate the $MN$ model to the 
$O(2)\times O(2)$ one, arguing that a fixed point with negative $\vb^*$
(located at $\ub^*\simeq 4.6$ and $\vb^*\simeq-4.0$ \cite{cops-02}) should 
exist, describing a finite temperature phase transition in the chiral 
universality class. 
Note that, as stressed in Refs. \cite{cp-01,cps-02,cops-02} (see also 
\cite{chiral3d} for the three-dimensional analogous) such fixed point
is in the region of non-Borel summability: in fact,  
$z^*=\vb^*/\ub^*\simeq -0.86<z_1=-0.833$, where $z_1=R_{MN}/R_N$ delimits
the region where the series are Borel summable (see Sec. \ref{sec2c}).
Anyway, in the course of the resummation with the conformal mapping, the 
singularity of the Borel transform closest to the origin has been taken into 
account. The resulting sequence of approximations is thus only asymptotic
(as for the random models), but it is expected to provide a reasonable 
estimate as long as the resummation point is far from the region 
$z_2=z_1 2N/(N+1) $, where the singularity on the real positive axis becomes
the closest to the origin. 
%%Thus for $z<z_2$ the resummation with the 
%%conformal mapping makes no sense anymore.

With this in mind, we resum the $\beta$ functions for several values of $M$
and $N$ and search for new fixed points with $\vb^*<0$.
We consider for each $\beta$ functions the 18 conformal mapping approximants
with $\alpha=0,1,2$ and $b=5,7,9,11,13,15$ that (as for the $O(M)\times O(N)$
model \cite{cops-02}) turned out to be the more stable with changing the 
number of terms considered in perturbation theory.
Up to three loops we do not find a fixed point for any value
of $N$ and $M$, but at four and five loops a fixed point with $z^*$ very 
close to $z_1$ appears for the majority of the 324 couples of approximants
of the $\beta$ functions. In Table \ref{tabMN} we report the five-loop results
for the location of the fixed point $\ub^*,\vb^*$ and the percentage of the 
324 approximant having it. In the Table are reported also the values of $z^*$ 
(obtained independently in the course of the averaging procedure) 
and the values of $z_1$ and $z_2$ to make the comparison between them 
clear at first sight.
The reported estimates are the averages over those approximants having a 
fixed point and the error bars are the variances.

\begin{table}[t]
\caption{Location $(\ub^*,\vb^*)$ of the stable fixed point with $\vb^*<0$ 
at five loops for several $M$ and $N$. 
The estimate of $z^*=\vb^*/\ub^*$ and the value of $z_1$ and $z_2$ (see 
the text) are also reported.}
\label{tabMN}
\begin{tabular}{cc|cccccc}
M  & N & \% &  $\ub^*$ &  $\vb^*$  &  $z^*$   & $z_1$ & $z_2$ \\
\tableline \hline
2  & 2 & 82 & 4.65(18) & -3.93(33) & -0.85(5) & -0.83 & -1.11\\
2  & 3 & 57 & 4.66(41) & -3.73(46) & -0.80(5) & -0.71 & -1.07\\
2  & 4 & 42 & 4.88(49) & -3.72(51) & -0.76(5) & -0.62 & -1\\
2  & 5 & 31 & 4.99(51) & -3.64(44) & -0.73(5) & -0.56 & -0.93\\
3  & 2 & 86 & 5.37(25) & -4.34(29) & -0.81(4) & -0.79 & -1.05\\
3  & 3 & 65 & 5.20(60) & -3.89(53) & -0.75(5) & -0.65 & -0.97\\
3  & 4 & 44 & 5.34(62) & -3.79(54) & -0.71(5) & -0.55 & -0.88\\
3  & 5 & 36 & 5.81(82) & -3.95(65) & -0.68(5) & -0.48 & -0.80\\
4  & 2 & 89 & 6.04(39) & -4.71(26) & -0.78(4) & -0.75 & -1\\
4  & 3 & 67 & 5.64(79) & -4.01(61) & -0.71(5) & -0.6  & -0.9\\
4  & 4 & 52 & 6.03(97) & -4.05(76) & -0.67(5) & -0.5  & -0.8\\
5  & 2 & 91 & 6.63(57) & -5.05(31) & -0.76(4) & -0.72 & -0.96\\
5  & 3 & 67 & 5.94(89) & -4.07(58) & -0.69(5) & -0.56 & -0.85\\
5  & 4 & 50 & 6.5(1.2) & -4.15(77) & -0.64(4) & -0.46 & -0.74\\
6  & 2 & 90 & 7.10(75) & -5.32(40) & -0.75(4) & -0.7  & -0.93\\
6  & 3 & 67 & 6.2(1.0) & -4.13(61) & -0.67(5) & -0.53 & -0.81\\
7  & 2 & 93 & 7.5(1.0) & -5.55(51) & -0.74(4) & -0.68 & -0.91\\
8  & 2 & 88 & 7.7(1.1) & -5.68(66) & -0.74(3) & -0.67 & -0.89\\
8  & 3 & 64 & 6.5(1.2) & -4.2(6)   & -0.58(2) & -0.5  & -0.75\\
9  & 2 & 84 & 7.9(1.2) & -5.83(70) & -0.74(3) & -0.65 & -0.87\\
10 & 2 & 83 & 8.0(1.2) & -5.87(70) & -0.74(3) & -0.64 & -0.86\\
15 & 2 & 75 & 8.1(1.0) & -6.01(55) & -0.74(3) & -0.61 & -0.81\\
15 & 4 & 15 & 10(2)    & -5.2(9)   & -0.52(1) & -0.34 & -0.54\\
20 & 2 & 55 & 8.9(1.0) & -6.57(54) & -0.74(2) & -0.58 & -0.78\\
20 & 3 & 55 & 8(1)     & -4.6(5)   & -0.58(2) & -0.41 & -0.62\\
22 & 2 & 23 & 9.2(5)   & -6.7(3)   & -0.73(2) & -0.58 & -0.77\\
25 & 2 & 21 & 10(1)    & -7.5(5)   & -0.73(2) & -0.57 & -0.76\\
\end{tabular}
\end{table}

For $M=N=2$ we reproduce the result of the chiral model \cite{cops-02}.
Several features for other value of $M$ and $N$ can be extracted from
Table \ref{tabMN}. For each M there is a maximum $N$, we call $N^*(M)$,
for which the FP exists. This is true up to a maximum value of $M$, 
we call $M^*$, after that the results of the $\e$ expansion and large $M$ 
are recovered. 
A reliable determination of $M^*$ and $N^*(M)$ is difficult.
In fact, one has arbitrarily to fix a confidence level of the percentage
for which the existence of the fixed point can be considered firm.
For example, we can decide that the fixed point is credible when the 
percentage of the approximants displaying it is greater than the $70\%$ and
its existence is improbable when it is less than $40\%$.
Within this method we have $2<N^*(2)<4$, $3<N^*(3)<5$, etc.
Regarding $M^*$ the situation is even more difficult. With increasing $M$,
the FP moves closer and closer to $z_2$, and, in fact, its existence is not
really reliable. Anyway, we retain quite safe to state $M^*<22$, but 
probably its value is much lower.

We check the stability of the results at four and five loops. For small
values of $M$ the location of the FP is rather stable, but it gets worse 
with increasing $M$, signaling that the estimate in this case is not so 
robust.

We evaluated the eigenvalues of the $\Omega$ matrix for each couple of 
approximants at their common zero. They always have a quite large positive 
real part, so the FP is stable. However, it is very hard to obtain reliable 
numerical estimates since the values strongly oscillate. At the same time, 
the focus behavior seems to be a peculiarity of the chiral model $M=N=2$; 
for larger $M,N$ only few approximants lead to a non-vanishing imaginary part. 
Even the estimates of the critical exponents are practically impossible 
probably because of non-analyticities and the large values of the coupling 
constants involved in the calculation. Anyway, the exponent $\eta$ for 
moderate $M$ seems to be compatible with $1/4$ that is the value of the 
Ising and $XY$ models.

Let us comment that a parallel analysis in $d=3$ \cite{cp-prep} 
seems to indicate a similar structure of fixed points, but the 
results are more stable (as usual) both because of the knowledge of
six-loop terms \cite{pv-00} and because of the weaker effect of 
non-analyticities (in particular, for the estimates of the exponents).

To conclude this section, we want to stress that Table \ref{tabMN} should be 
read with care. The point is that all the found FP have $z^*<z_1$ being located  
in the region where the series are not Borel summable and all the FP 
with $z^*$ close to $z_2$ are not credible because of the bad behavior 
of the resummed approximants close to this region.

\section{Conclusion}
\label{sec4}

To summarize, the critical behaviors of the two-dimensional $N$-vector cubic 
and $MN$ models have been studied within the renormalization-group approach. 
The five-loop contributions to the $\beta$ functions and critical 
exponents have been calculated and the five-loop RG series have been resummed 
by means of Pad\'e-Borel-Leroy procedure and the conformal mapping technique.
 
For the cubic planar model ($N = 2$) we have found that the continuous line 
of fixed points connecting  the Heisenberg and the Ising ones is well 
reproduced by the resummed five-loop RG series. 
Moreover, the five-loop terms being taken into account make the lines of 
zeros of $\beta$ functions for $\ub$ and $\vb$ closer to each another 
thus improving the results of the lower-order approximation. 
For the cubic model with $N>2$, the five-loop contributions have been shown 
to shift the cubic fixed point, given by the four-loop approximation, towards
the Ising one. 
This may be considered as an argument in favor of the idea that the existence 
of cubic fixed point in two dimensions for $N\geq 3$ is an artifact of 
the perturbative analysis. 

The models with $N=0$ describing the critical thermodynamics 
of two-dimensional weakly-disordered $O(M)$ systems has been also studied. 
The results obtained have been found to be compatible with the conclusion 
that in two dimensions the impure critical behavior is governed by the 
$O(M)$ fixed point, even in the Ising case, where $\alpha_I=0$ and the Harris
criterion is inconclusive. 

The five-loop RG analysis of the two-dimensional $MN$-model with $M,N\geq2$ 
has been also performed. For $v_0>0$ we reproduced all the non-perturbative 
results. It was shown, in particular, that the transition is driven to the 
$O(M)$ fixed point, that only for $M\leq2$ describes a finite temperature 
phase transition. 
We also found a stable fixed point in the region with $v_0<0$ that has no 
counterpart in $\e$ and large $M$ expansion, but its location is in the 
region of non-Borel summability of the series and its existence is still 
doubtful.
At fixed $M$, this new fixed point is found for $N<N^*(M)$ up to a maximum
value $M^*$, after which it disappears. Whether this fixed point describes a 
finite temperature phase transition (allowed because of the discrete 
symmetry $C_N$) or a zero temperature one cannot be discerned by 
our analysis; only lattice techniques such as Monte Carlo simulation and high 
temperature expansion or real experiments can completely clarify this point.

\section*{Acknowledgments}

We are grateful to Pietro Parruccini and Ettore Vicari for discussions.
The authors acknowledge the financial support of 
the Russian Foundation for Basic Research under Grant 
No. 04-02-16189 (A.I.S., E.V.O., D.V.P.), and the Ministry 
of Education of Russian Federation under Grants 
No. E02-3.2-266 (A.I.S., E.V.O., D.V.P.),
No. A03-2.9-227 (A.I.S., D.V.P.),
and of EPSRC under Grant No. GR/R83712/01 (P.C.).
A.I.S. has much benefited 
from the warm hospitality of Scuola Normale Superiore and 
Dipartimento di Fisica dell'Universit\'a di Pisa, where 
the major part of this research was done.


\begin{references}
\bibitem{OS-corr} Unfortunately, the five-loop contribution to 
the critical exponent $\gamma$ ($\eta_t$) reported in Ref. \cite{OS} 
contains numerical errors. Here, the correct values of 
corresponding coefficients are presented (Table \ref{t4}, line [5,0]).  
This correction improves a bit the estimate of the exponent $\nu$ for the 
two-dimensional Ising model. Indeed, applying the same method of 
Ref. \cite{OS} (i.e. the Pad\'e-Borel [4/1] with $b=1$ to the series 
$1/\nu(g)$), we obtain at $g^*=1.837$ the estimate $\nu=1.017$, that is
closer to the exact value $\nu=1$.
\bibitem{pseudo}
The pseudo-$\e$ expansion was introduced by B. Nickel, see the second 
of Ref.\cite{LZ-77}.
\end{references}
\end{document}